\documentclass[aps,prd,reprint,superscriptaddress,nofootinbib]{revtex4-2}

\usepackage[T1]{fontenc}
\usepackage{amsmath,amssymb,amsfonts,mathtools}
\usepackage{bm}
\usepackage{graphicx}
\usepackage{siunitx}
\usepackage{microtype}

\allowdisplaybreaks
\DeclareSIUnit\dalton{Da}
\DeclareSIUnit\parsec{pc}

\newcommand{\A}{\mathcal{A}}
\newcommand{\Tr}{\operatorname{Tr}}
\newcommand{\dd}{\mathrm{d}}
\newcommand{\ii}{\mathrm{i}}
\newcommand{\ee}{\mathrm{e}}
\newcommand{\eps}{\epsilon}
\newcommand{\Mpl}{m_{\mathrm P}}
\newcommand{\order}{\mathcal{O}}

\usepackage{color}
\definecolor{phthaloblue}{rgb}{0.0, 0.06, 0.54}
\definecolor{purple}{rgb}{0.5 ,0, 0.7}
\definecolor{bluegreen}{rgb}{0, 0.45, 0.35}
\definecolor{sakura}{rgb}{1 ,0.52, 0.74}
\definecolor{wakakusa}{rgb}{0.45 ,0.74, 0}
\definecolor{brown}{rgb}{0.48 ,0.23, 0}
\definecolor{skyblue}{rgb}{0.21 ,0.7, 1.}
\definecolor{purplegray}{rgb}{0.35,0.35,0.73}
\usepackage{hyperref}
\hypersetup{colorlinks=true, linkcolor=bluegreen, citecolor=purplegray, urlcolor=purplegray}

\begin{document}

\title{Gravitational decoherence from soft-graviton \texorpdfstring{$S$}{S}-matrix}
\author{Hiroki Matsui}
\affiliation{Osaka Central Advanced Mathematical Institute (OCAMI), Osaka Metropolitan University, Osaka 558-8585, Japan}
\date{\today}

\begin{abstract}
We revisit gravitational decoherence of matter superpositions using the
leading soft sector of the graviton $S$-matrix. Starting from Weinberg's
soft theorem and the overlap of coherent radiation states, we derive,
for branch configurations with the same total four-momentum, a
gauge-invariant, nonnegative leading-soft decoherence exponent and
evaluate the polarization sum and angular integrals exactly at this
order. In the vacuum, coherence between branches with distinguishable
soft radiation vanishes as the infrared cutoff is removed. The leading
nonrelativistic term is purely quadrupolar as a consequence
of energy--momentum conservation. We extend the analysis to single- and
two-mode squeezed graviton states, including the two-mode case motivated
by inflation. Squeezing can enhance or suppress decoherence depending on
the relative phase, while phase averaging gives an enhancement governed
by the graviton occupation. For a phase-averaged relic background with
an approximately flat energy-density spectrum over the relevant soft
frequency band, the sharp-cutoff identification of the infrared scale
with the inverse observation time gives a contribution that grows with
the gravitational-wave energy density and with the fourth power of the
observation time.
These results connect soft-graviton scattering, infrared quantum
information, and decoherence in nonvacuum gravitational backgrounds.
\end{abstract}

\maketitle

\section{Introduction}
\label{sec:intro}

Any interference experiment performed with a massive object is, unavoidably, an experiment coupled to the gravitational field. Because the graviton is massless, every scattering or branching process in which the matter degrees of freedom end up in distinct momentum configurations is accompanied by the emission of arbitrarily soft gravitational bremsstrahlung. The mean number of emitted soft gravitons diverges logarithmically in the infrared (IR), a fact known since Weinberg's classic analysis of infrared photons and gravitons~\cite{Weinberg:1965nx}. In inclusive rates this divergence famously cancels against virtual corrections~\cite{Weinberg:1965nx,Bloch:1937pw,Yennie:1961ad}. At the level of 
\emph{quantum information}, however, the soft radiation does not cancel: the soft gravitons emitted by two different branches of a matter superposition are macroscopically distinct field configurations, and tracing over them suppresses the off-diagonal elements of the matter density matrix. This mechanism---decoherence by soft emission---has been analyzed from several complementary viewpoints: inclusive Fock-space and asymptotic-radiation treatments~\cite{Calucci:2003vx,Carney:2017jut,DeLisle:2019dyw}, infrared-dressing formulations developed most explicitly in QED and perturbative gravity~\cite{Carney:2017oxp,Carney:2018softdressing,Chung:1965zza,Kibble:1968oo,Kibble:1968ms,Kulish:1970ut,Neuenfeld:2018fdw,Semenoff:2019dqe,Ware:2013zja}, asymptotic symmetries and memory~\cite{Strominger:2017zoo,Wilson-Gerow:2018egh,Linton:2025dcz}, and open-system master equations~\cite{Blencowe:2012mp,Anastopoulos:2013zya,Kanno:2020usf,Moreira:2026gdc}; see also Ref.~\cite{Hsiang:2024gpt} for a recent review. Related soft-radiation decoherence occurs in the presence of Killing horizons~\cite{Danielson:2022tdw,Danielson:2022sga,Gralla:2023oya,Wilson-Gerow:2024ljx,Danielson:2025softQI}.

The purpose of this paper is twofold. First, we present a self-contained and 
explicit derivation of gravitational decoherence directly from the soft-graviton $S$-matrix. 
The final result for the suppression of coherence between two momentum branches $\alpha$ and $\beta$ is
\begin{align}
|\rho_{\alpha\beta}|
&\longrightarrow |\rho_{\alpha\beta}|\,
\exp\!\left[-\frac{\Delta B_{\alpha\beta}}{2}\ln\frac{\Lambda}{\lambda}\right]
\nonumber\\
&=|\rho_{\alpha\beta}|
\left(\frac{\lambda}{\Lambda}\right)^{\Delta B_{\alpha\beta}/2},
\label{eq:intro-main}
\end{align}
where $\lambda$ and $\Lambda$ are the IR and soft ultraviolet (UV) ends of the graviton frequency integral and $\Delta B_{\alpha\beta}\ge 0$ is the Weinberg $B$-factor evaluated on the \emph{difference current} of the two branches, Eq.~\eqref{eq:DeltaB}. Coherence between distinct momentum configurations is therefore destroyed completely as $\lambda\to 0$: the soft gravitons carry away perfect which-path information in the strict IR limit~\cite{Carney:2017jut}. In the present work we rederive this result in terms of branch-conditioned
coherent radiation states, a formulation that isolates the decoherence exponent
as an overlap of field displacements and thereby provides the starting point for
the generalizations that follow. We also evaluate the nonrelativistic limit exactly, where energy and momentum
conservation between the branches cancels the monopole and dipole records and
leaves a manifestly quadrupolar result, Eq.~\eqref{eq:NRfinal}.

Second, we generalize the initial graviton state from the vacuum to
squeezed states, motivated in particular by the two-mode
squeezing generated during inflation~\cite{Grishchuk:1990bj,Albrecht:1992kf}.
The decoherence functional probes the graviton-state characteristic
function at the difference between the branch displacements. Squeezing
therefore makes decoherence quadrature dependent: the relative phase can
suppress or enhance it, whereas phase averaging leaves an enhancement
controlled by the graviton occupation number. For a flat relic spectrum,
the sharp-cutoff prescription considered below gives a contribution that
grows with the fourth power of the observation time. Squeezed graviton
states and their observational signatures have been discussed in
Refs.~\cite{Parikh:2020kfh,Parikh:2020fhy,Kanno:2020usf,Kanno:2021gpt,
Matsui:2025sqg,Miyauchi:2026cnt}. We also distinguish infrared-finite
dressed scattering amplitudes from the coherence of a reduced state
obtained after unresolved soft modes are traced out.

The present analysis is complementary to the finite-time treatment of matter-wave interferometric histories in Ref.~\cite{Matsui:2026whichpath}, which also considered vacuum and squeezed initial graviton states. While that work treats the driven linearized-gravity evolution for branch differences localized in time, the present work focuses on the universal leading-soft infrared structure of asymptotically distinct momentum branches. In both approaches, decoherence is governed by the overlap of branch-dependent graviton displacements, providing a direct connection between the finite-time and soft-$S$-matrix descriptions.

The paper is organized as follows. Section~\ref{sec:conventions} fixes conventions and reviews the soft-graviton theorem and its eikonal exponentiation into coherent radiation states. Section~\ref{sec:decoherence} derives the decoherence functional from the reduced density matrix. Section~\ref{sec:evaluation} evaluates the polarization sum, the angular integral, and the frequency integral in full detail, establishes positivity, and states the main result. Section~\ref{sec:NR} performs the nonrelativistic expansion and derives the quadrupole formula. Section~\ref{sec:squeezed} generalizes to single-mode and two-mode squeezed graviton states and derives the modified decoherence laws, including the relic-background estimate. Section~\ref{sec:discussion} discusses the physical implications, orders of magnitude, and limitations, including the relation to infrared-finite dressed-state descriptions. Four appendices collect the derivations of the graviton polarization sum (Appendix~\ref{app:polsum}), the master angular integral (Appendix~\ref{app:angular}), the displacement- and squeeze-operator identities (Appendix~\ref{app:squeeze}), and the details of the nonrelativistic expansion (Appendix~\ref{app:NR}).

\paragraph*{Conventions.}
We use natural units $\hbar=c=1$ except where indicated and the mostly-plus metric
\begin{equation}
\eta_{\mu\nu}=\operatorname{diag}(-1,+1,+1,+1).
\label{eq:metric}
\end{equation}
Our Fourier convention is
\begin{align}
\widetilde{F}(\omega,\bm q)
&=\int\!\dd^4x\,\ee^{\ii\omega t-\ii\bm q\cdot\bm x}F(t,\bm x)
\nonumber\\
&=\int\!\dd^4x\,\ee^{-\ii q\cdot x}F(x),
\label{eq:FourierConvention}
\end{align}
where $q^\mu=(\omega,\bm q)$.
We define
\begin{equation}
g_{\mu\nu}=\eta_{\mu\nu}+\kappa\, h_{\mu\nu},\qquad
\kappa\equiv\sqrt{32\pi G},\qquad \Mpl\equiv G^{-1/2}.
\label{eq:kappa}
\end{equation}

\section{Conventions, soft theorem, and eikonal exponentiation}
\label{sec:conventions}

\subsection{Graviton mode expansion}

Expanding the Einstein--Hilbert action to quadratic order in $h_{\mu\nu}$ and coupling to matter through
\begin{equation}
\mathcal{L}_{\rm int}=+\frac{\kappa}{2}\,h_{\mu\nu}T^{\mu\nu},
\label{eq:Lint}
\end{equation}
the free graviton field in the interaction picture has the mode expansion
\begin{equation}
h_{\mu\nu}(x)=\sum_{h=\pm 2}\int\!\frac{\dd^{3}q}{(2\pi)^{3}}\frac{1}{\sqrt{2\omega_{q}}}
\Big[\eps^{h}_{\mu\nu}(\bm q)\,a_{h}(\bm q)\,\ee^{+\ii q\cdot x}+\text{h.c.}\Big],
\label{eq:modeexp}
\end{equation}
with $\omega_q=|\bm q|$, canonical commutators
\begin{equation}
[a_{h}(\bm q),a^{\dagger}_{h'}(\bm q')]=(2\pi)^{3}\delta_{hh'}\,\delta^{3}(\bm q-\bm q'),
\end{equation}
and transverse-traceless polarization tensors normalized as
\begin{equation}
\eps^{h}_{\mu\nu}\eps^{h'*\,\mu\nu}=\delta^{hh'},\qquad
q^{\mu}\eps^{h}_{\mu\nu}(\bm q)=0,\qquad \eps^{h\,\mu}{}_{\mu}=0 .
\label{eq:polnorm}
\end{equation}
It is convenient to build $\eps^{\pm2}_{\mu\nu}=\eps^{\pm}_{\mu}\eps^{\pm}_{\nu}$ from photon helicity vectors $\eps^{\pm}_{\mu}(\bm q)$; the completeness relation on the physical subspace is derived in Appendix~\ref{app:polsum}. We adopt the parity convention
\begin{equation}
\eps^{h}_{\mu\nu}(-\bm q)=\eps^{h*}_{\mu\nu}(\bm q),
\label{eq:parity}
\end{equation}
which is convenient for the two-mode squeezing analysis of Sec.~\ref{sec:squeezed} and is consistent with the reality of $h_{\mu\nu}(x)$.

\subsection{Weinberg soft factor}

Consider a scattering process from a fixed incoming state to a final
configuration $\alpha=\{p_i\}_{i\in\alpha}$, and denote its amplitude by
$M_\alpha$. The label $\alpha$ specifies the complete outgoing
configuration, whereas $i$ labels an individual particle; spin and
species labels are suppressed. We write ${\rm in}$ for the set of
incoming particles. All hard momenta $p_i^\mu=(E_i,\bm p_i)$ are physical,
future-directed momenta, including those of the incoming particles.

We compare this process with one in which an additional outgoing
graviton is emitted, with four-momentum
\begin{equation}
q^\mu=(\omega_q,\bm q),
\qquad \omega_q=|\bm q|,
\qquad q^2=0,
\end{equation}
and helicity $h=\pm2$. The \emph{soft limit} means
$\omega_q\to0$ at fixed emission direction
$\hat{\bm q}=\bm q/\omega_q$, with $\omega_q$ much smaller than every
relevant energy or momentum-transfer scale of the underlying process.
The word \emph{hard} is relative to this soft scale and the hard amplitude on the
right-hand side of the soft theorem is evaluated at the limiting
momenta obtained as $q\to0$.

The leading soft singularity arises from graviton emission off an external
incoming or outgoing leg. For an external particle of momentum $p_i$, the
adjacent propagator carries momentum $p_i\pm q$, with the upper (lower)
sign for an outgoing (incoming) leg. Using $p_i^2=-m_i^2$ and $q^2=0$,
\begin{equation}
\frac{1}{(p_i\pm q)^2+m_i^2}
=\frac{1}{\pm2p_i\cdot q}.
\end{equation}
Since $p_i\cdot q\propto\omega_q$, this produces the characteristic
$1/\omega_q$ soft pole. The eikonal approximation retains this pole while
evaluating the leading gravitational vertex, proportional to
$p_i^\mu p_i^\nu$, and the hard amplitude at $q=0$. Recoil corrections
and emission from generic off-shell internal lines contribute only at
subleading order.

Combining the vertex and propagator factors and summing over all
external particles gives Weinberg's soft theorem
\cite{Weinberg:1965nx},
\begin{equation}
M_{\alpha+g(q,h)}
=\frac{\kappa}{2}\,
\eps^{h*}_{\mu\nu}(\bm q)\,
J^{\mu\nu}_\alpha(q)\,M_\alpha
+\order(\omega_q^0),
\label{eq:soft}
\end{equation}
where 
\begin{equation}
J^{\mu\nu}_{\alpha}(q)
=\sum_{i\in\alpha\,\cup\,{\rm in}}\eta_i\,
\frac{p_i^\mu p_i^\nu}{p_i\cdot q},
\qquad
\eta_i=
\begin{cases}
+1 & \text{outgoing},\\
-1 & \text{incoming}.
\end{cases}
\label{eq:eikonalcurrent}
\end{equation}
The complex-conjugated polarization tensor is appropriate to an
emitted graviton in our mode-expansion convention. The tensor
$J^{\mu\nu}_\alpha$ is a momentum-space soft current encoding the
asymptotic incoming and outgoing motion, rather than a local
stress--energy tensor. The factor multiplying $M_\alpha$ is universal:
at this order it depends on the external momenta and the gravitational
coupling, but not on the detailed interaction producing $\alpha$ or on
the spins of the hard particles.

For a future-directed massive particle in the mostly-plus convention,
\[
p_i\cdot q
=-\omega_q\bigl(E_i-\bm p_i\cdot\hat{\bm q}\bigr)<0,
\]
since $E_i>|\bm p_i|$. The factor $\eta_i=\pm1$ simply distinguishes
outgoing and incoming legs and is specific to the asymptotic $S$-matrix
description; it is not required in the finite-time formulation of
Ref.~\cite{Matsui:2026whichpath}.
Eq.~\eqref{eq:eikonalcurrent} therefore represents the difference
between outgoing and incoming soft currents. Contributions with unchanged
incoming and outgoing momenta cancel, and if the complete momentum
configuration is unchanged, the leading $1/\omega_q$ soft pole vanishes.
This does not preclude radiation at finite frequency.

In the soft limit, momentum conservation of the hard process implies
\begin{equation}
\sum_{i\in\alpha}p_i^\nu
-\sum_{i\in{\rm in}}p_i^\nu=0,
\end{equation}
and hence
\begin{equation}
q_\mu J^{\mu\nu}_\alpha(q)
=\sum_i\eta_i p_i^\nu
=0.
\label{eq:ward}
\end{equation}
This Ward identity ensures gauge invariance of the leading soft factor.
At finite $q$, momentum conservation also includes the emitted graviton;
the resulting recoil corrections enter only at subleading order.

Under a gauge transformation
\begin{equation}
\epsilon^{h*}_{\mu\nu}
\rightarrow
\epsilon^{h*}_{\mu\nu}+q_\mu\xi_\nu+q_\nu\xi_\mu,
\end{equation}
Eq.~\eqref{eq:ward} gives
\begin{equation}
\delta_\xi\!\left(
\epsilon^{h*}_{\mu\nu}J^{\mu\nu}_\alpha
\right)
=2\xi_\nu q_\mu J^{\mu\nu}_\alpha=0,
\end{equation}
so the leading soft factor is gauge invariant. This requires the complete conserved set of external particles; in interferometric applications, any recoiling apparatus or partner needed for momentum conservation must therefore be included.

At tree level, the $\order(\omega_q^0)$ remainder is finite as
$\omega_q\to0$, and the first subleading soft operator involves each
hard particle's total angular momentum, including orbital and spin
contributions~\cite{Cachazo:2014fwa}. The leading soft factor,
\begin{equation}
S^{(0)}_{\alpha h}(q)
\equiv \frac{\kappa}{2}
\eps^{h*}_{\mu\nu}(\bm q)J^{\mu\nu}_\alpha(q)
\propto \omega_q^{-1},
\end{equation}
combined with the four-dimensional phase-space measure,
\begin{equation}
\frac{\dd^3q}{(2\pi)^3\,2\omega_q}
=\frac{\omega_q\,\dd\omega_q\,\dd\Omega_{\hat{\bm q}}}
{2(2\pi)^3},
\end{equation}
produces the vacuum infrared logarithm
\begin{equation}
\int_\lambda^\Lambda \omega_q\,\dd\omega_q\,
\bigl|S^{(0)}_{\alpha h}(q)\bigr|^2
\propto
\int_\lambda^\Lambda\frac{\dd\omega_q}{\omega_q}
=\ln\frac{\Lambda}{\lambda},
\end{equation}
with angular factors suppressed. Here $\lambda>0$ is an infrared
regulator, and $\Lambda$ lies below all relevant hard scales.
Leading--subleading interference and the square of the finite
remainder instead scale as $\int\dd\omega_q$ and
$\int\omega_q\,\dd\omega_q$, respectively, and are infrared finite
in vacuum.

The same counting applies to the soft-factor difference between two
hard branches and hence to the vacuum decoherence logarithm.
Subleading terms give finite corrections and may dominate when the
leading difference vanishes. In nonvacuum backgrounds, occupation
and squeezing modify the infrared weighting. Retaining only the
leading soft term therefore requires the omitted state-weighted
contributions to be subdominant; they need not be infrared finite
for an arbitrary initial graviton state.

\subsection{Exponentiation into coherent states}

The leading soft theorem has an important consequence when more than one
soft graviton is emitted. At leading soft order, the emission of multiple gravitons factorizes into
a product of single-graviton soft factors. For $n$ gravitons with momenta
$q_a$ and helicities $h_a$,
\begin{equation}
M_{\alpha+n g}
\simeq
M_\alpha
\prod_{a=1}^{n}
\left[
\frac{\kappa}{2}
\eps^{h_a*}_{\mu\nu}(\bm q_a)
J^{\mu\nu}_\alpha(q_a)
\right],
\end{equation}
up to terms suppressed by powers of the soft energies relative to the
hard scales of the process.
This factorization leads to the exponentiation of soft radiation~\cite{Weinberg:1965nx,Chung:1965zza,Kibble:1968oo,Kibble:1968ms}.

Summing over arbitrary numbers of emitted soft gravitons with the
appropriate Bose symmetry factors exponentiates the soft emission into a
displacement operator. Conditioned
on a hard final branch $\alpha$, the outgoing state takes the form
\begin{align}
S\,|i\rangle\otimes|0\rangle_{g}
&=\sum_{\alpha} S_{\alpha i}\;|\alpha\rangle\otimes \ee^{\ii\Phi_{\alpha}}\,\hat D[\A_{\alpha}]\,|0\rangle_{g},
\label{eq:branching}\\
\hat D[\A_{\alpha}]&=\exp\!\Big[\sum_{h}\int\!\frac{\dd^{3}q}{(2\pi)^3}
\big(\A^{\alpha}_{h}(\bm q)\,a^{\dagger}_{h}(\bm q)-\text{h.c.}\big)\Big],
\label{eq:displacement}
\end{align}
Here $S_{\alpha i}$ is the hard scattering amplitude, while
$\hat D[\A_\alpha]$ generates the branch-dependent soft-graviton coherent
state. The branch-dependent displacement amplitude is fixed directly by the
leading soft factor,
\begin{equation}
\A^\alpha_h(\bm q)
=
\frac{\kappa}{2}
\frac{
\eps^{h*}_{\mu\nu}(\bm q)
J^{\mu\nu}_\alpha(q)}
{\sqrt{2\omega_q}} .
\label{eq:Adef}
\end{equation}

The connection with multiple soft emission becomes transparent by
normal-ordering the displacement operator. Defining
\begin{equation}
\|\A_\alpha\|^2
\equiv
\sum_h\int\frac{\dd^3q}{(2\pi)^3}
|\A^\alpha_h(\bm q)|^2 ,
\end{equation}
we obtain
\begin{align}
\hat D[\A_\alpha]|0\rangle_g
&=
\ee^{-\|\A_\alpha\|^2/2}
\nonumber\\[-2pt]
&\quad\times
\exp\!\left[
\sum_h\int\frac{\dd^3q}{(2\pi)^3}
\A^\alpha_h(\bm q)a_h^\dagger(\bm q)
\right]
|0\rangle_g .
\end{align}
Expanding the exponential generates the
$n$-graviton components with their Bose symmetry factors, the prefactor
supplying the no-emission normalization. Thus, the leading-soft radiation
state is a multimode coherent state,
\begin{equation}
|\gamma_\alpha\rangle
\equiv
\hat D[\A_\alpha]|0\rangle_g ,
\qquad
a_h(\bm q)|\gamma_\alpha\rangle
=
\A^\alpha_h(\bm q)|\gamma_\alpha\rangle .
\end{equation}
This has a simple physical interpretation. A coherent state is the
quantum state that most closely resembles a classical radiation field:
its expectation value of the graviton field follows the classical
linearized bremsstrahlung solution. The function
$\A^\alpha_h(\bm q)$ therefore specifies the positive-frequency mode
content of the soft gravitational radiation associated with the hard
branch $\alpha$. Different hard momentum configurations generally
produce different coherent radiation states, and their distinguishability
will be the origin of decoherence in Sec.~\ref{sec:decoherence}.
Also, we denote this current by $J^{\mu\nu}$ to distinguish it from
the spacetime stress tensor $T^{\mu\nu}(x)$ and its Fourier transform $\widetilde T^{\mu\nu}$ 
used in Ref.~\cite{Matsui:2026whichpath}.

The additional phase $\Phi_\alpha$ arises from virtual soft
graviton exchange among the external legs of branch $\alpha$, often
referred to as the gravitational Coulomb or Weinberg phase~\cite{Weinberg:1965nx}.
 It is diagonal in the hard branch and affects
the relative phase of interference amplitudes, but not directly the
magnitude of the coherent-state overlap considered below.

\section{Reduced density matrix and the decoherence functional}
\label{sec:decoherence}

We now consider an initial state which branches into 
a superposition of two hard configurations,
\begin{equation}
|\psi\rangle
=c_{\alpha}|\alpha\rangle+c_{\beta}|\beta\rangle,
\qquad
|c_{\alpha}|^{2}+|c_{\beta}|^{2}=1,
\end{equation}
where $\alpha$ and $\beta$ denote distinct momentum configurations,
and $c_{\alpha}$ and $c_{\beta}$ are the corresponding normalized
branch amplitudes. We assume that the two configurations carry the
same conserved total four-momentum $P^{\mu}$.

At leading soft order, Eq.~\eqref{eq:branching} implies that the
interaction correlates each hard branch with its corresponding
branch-dependent coherent graviton state. The resulting joint
matter--graviton state is therefore
\begin{equation}
|\Psi\rangle
=c_{\alpha}\ee^{\ii\Phi_{\alpha}}
|\alpha\rangle\otimes|\gamma_{\alpha}\rangle
+c_{\beta}\ee^{\ii\Phi_{\beta}}
|\beta\rangle\otimes|\gamma_{\beta}\rangle .
\end{equation}
Here, $\Phi_{\alpha}$ and $\Phi_{\beta}$ are the branch-dependent
phases, while $|\gamma_{\alpha}\rangle$ and
$|\gamma_{\beta}\rangle$ describe the corresponding soft-graviton
radiation states.

The reduced density matrix of the matter sector is obtained by tracing
over the graviton degrees of freedom,
$\rho=\Tr_{g}|\Psi\rangle\langle\Psi|$.
Its off-diagonal matrix element between the two hard branches is
therefore
\begin{equation}
\rho_{\alpha\beta}
\equiv\langle\alpha|\rho|\beta\rangle
=c_{\alpha}c^{*}_{\beta}\,
\ee^{\ii(\Phi_{\alpha}-\Phi_{\beta})}\,
\langle\gamma_{\beta}|\gamma_{\alpha}\rangle .
\label{eq:rhooff}
\end{equation}
Thus, the overlap of the two radiation states directly controls the
survival of coherence between the hard branches; in particular, its
modulus determines the suppression of the corresponding off-diagonal
density-matrix element.

The standard inner product for multimode coherent states (see
Appendix~\ref{app:squeeze}) gives
\begin{equation}
\langle\gamma_{\beta}|\gamma_{\alpha}\rangle
=\exp\!\Big[\!\sum_h\int\!\frac{\dd^3q}{(2\pi)^3}\Big(
\A^{\beta*}_h\A^{\alpha}_h-\tfrac12|\A^{\alpha}_h|^{2}-\tfrac12|\A^{\beta}_h|^{2}\Big)\Big],
\end{equation}
Taking the modulus gives the suppression factor for the off-diagonal
coherence, which we express in terms of the \emph{decoherence functional}
$\Gamma_{\alpha\beta}$ as
\begin{equation}
|\langle\gamma_{\beta}|\gamma_{\alpha}\rangle|
=\ee^{-\Gamma_{\alpha\beta}},\qquad
\Gamma_{\alpha\beta}
=\frac{1}{2}\sum_{h}\int\!\frac{\dd^{3}q}{(2\pi)^3}
\big|\Delta\A_{h}(\bm q)\big|^{2},
\label{eq:Gammadef}
\end{equation}
where
$\Delta\A_h(\bm q)\equiv
\A^{\alpha}_h(\bm q)-\A^{\beta}_h(\bm q)$.
Equation~\eqref{eq:Gammadef} admits a direct physical interpretation.
Defining
\begin{equation}
N_{\Delta}
\equiv
\sum_h\int\!\frac{\dd^3q}{(2\pi)^3}
\big|\Delta\A_h(\bm q)\big|^{2},
\qquad
\Gamma_{\alpha\beta}=\tfrac{1}{2}N_{\Delta},
\label{eq:halfN}
\end{equation}
we may regard $N_{\Delta}$ as the mean graviton occupation number of
the coherent state associated with the relative displacement between
the two radiation clouds. Equivalently, it measures the mean number of
gravitons that can carry information distinguishing the two branches.
Thus, $N_{\Delta}=1$ reduces the overlap modulus, and hence the
off-diagonal coherence, by a factor of $\ee^{-1/2}$, whereas
$N_{\Delta}\gg1$ renders the residual coherence exponentially small
and the branches effectively decoherent.

Because the two branches originate from the same initial hard state,
their common incoming-leg contributions cancel in the relative
displacement. At leading soft order, $\Delta\A$ is therefore determined
by the \emph{difference current} of the two outgoing configurations:
\begin{align}
\Delta\A_{h}(\bm q)
&=\frac{\kappa}{2}\,
\frac{\eps^{h*}_{\mu\nu}(\bm q)\,
\Delta J_{\alpha\beta}^{\mu\nu}(q)}
{\sqrt{2\omega_{q}}},
\label{eq:DeltaA}\\
\Delta J_{\alpha\beta}^{\mu\nu}(q)
&\equiv
\sum_{i\in\alpha\cup\beta}\tilde\eta_{i}\,
\frac{p_{i}^{\mu}p_{i}^{\nu}}{p_{i}\cdot q},
\qquad
\tilde\eta_{i}=
\begin{cases}
+1, & i\in\alpha,\\
-1, & i\in\beta.
\end{cases}
\label{eq:diffcurrent}
\end{align}
Since the two outgoing configurations carry the same total
four-momentum,
$\sum_{i\in\alpha}p_i^\nu
=\sum_{j\in\beta}p_j^\nu=P^\nu$,
the difference current satisfies
\begin{equation}
q_{\mu}\Delta J_{\alpha\beta}^{\mu\nu}(q)
=\sum_{i\in\alpha\cup\beta}\tilde\eta_{i}\,p_{i}^{\nu}
=P^{\nu}-P^{\nu}=0.
\label{eq:diffward}
\end{equation}
Thus, $\Delta J_{\alpha\beta}^{\mu\nu}$ is conserved. This Ward identity
ensures the gauge invariance of $\Gamma_{\alpha\beta}$ and provides the
conservation condition underlying the multipole cancellations derived
in Sec.~\ref{sec:NR}. Its nonnegativity, by contrast, is manifest
directly from the norm-square expression in Eq.~\eqref{eq:Gammadef}.

An important physical implication concerns momentum superpositions.
Within the closed-system scattering description adopted here, retaining
only a single free particle ``in a superposition of two momenta'' while
omitting the degrees of freedom that carry the compensating recoil gives
an incomplete description of the branching process. The recoiling
system---for example, a beam splitter, apparatus, or emitter---must also
be included. Its contributions enter
$\Delta J_{\alpha\beta}^{\mu\nu}$ at the same leading soft order and are
required for Eq.~\eqref{eq:diffward} to hold. If these contributions are
omitted, the resulting truncated current is not conserved, and the
calculated decoherence can acquire gauge artifacts and spurious,
potentially unbounded behavior.

At leading soft order, the coherent-state overlap contributes no phase:
Eqs.~\eqref{eq:DeltaA} and \eqref{eq:polsumresult} express
$\sum_h\A^{\beta*}_h\A^\alpha_h$ as a real contraction of the branch
currents. The relative phase therefore arises solely from the difference
$\Phi_\alpha-\Phi_\beta$ of Weinberg's virtual-exchange phases,
\begin{equation}
\Phi_X
=\frac{G}{2}\sum_{i\neq j\in X}
\eta_i\eta_j m_i m_j
\frac{1+\beta_{ij}^{2}}
{\beta_{ij}\sqrt{1-\beta_{ij}^{2}}}
\ln\frac{\Lambda}{\lambda}
+\cdots ,
\end{equation}
which encodes the eikonal gravitational interaction among the external
legs~\cite{Weinberg:1965nx}. This phase shifts the interference fringes
but does not affect their visibility $|\rho_{\alpha\beta}|$; we therefore
do not discuss it further.

\section{Evaluation of the decoherence exponent}
\label{sec:evaluation}

We now evaluate Eq.~\eqref{eq:Gammadef} explicitly within the
leading-soft approximation. Substituting Eq.~\eqref{eq:DeltaA}, we obtain
\begin{equation}
\Gamma_{\alpha\beta}
=\frac{\kappa^{2}}{8}
\int\!\frac{\dd^{3}q}{(2\pi)^{3}\,2\omega_{q}}
\sum_{h}
\big|
\eps^{h*}_{\mu\nu}(\bm q)\,
\Delta J_{\alpha\beta}^{\mu\nu}(q)
\big|^{2}.
\label{eq:Gamma1}
\end{equation}
The computation splits into a polarization sum, an angular integral, and a frequency integral.

\subsection{Polarization sum}
\label{sec:polsum}

For a null momentum $q^\mu=\omega(1,\hat{\bm q})$, we introduce the
auxiliary null vector
$\bar q^\mu=\omega(1,-\hat{\bm q})$, which satisfies
$q\cdot\bar q=-2\omega^2$. The projector onto the subspace transverse
to both $q^\mu$ and $\bar q^\mu$ is
\begin{equation}
\Pi_{\mu\nu}
=\eta_{\mu\nu}
-\frac{q_{\mu}\bar q_{\nu}+\bar q_{\mu}q_{\nu}}
{q\cdot\bar q}.
\end{equation}
The sum over the two physical graviton helicities is then
(Appendix~\ref{app:polsum})
\begin{equation}
\sum_{h=\pm2}
\eps^{h}_{\mu\nu}(\bm q)\,
\eps^{h*}_{\rho\sigma}(\bm q)
=\frac{1}{2}
\Big(
\Pi_{\mu\rho}\Pi_{\nu\sigma}
+\Pi_{\mu\sigma}\Pi_{\nu\rho}
-\Pi_{\mu\nu}\Pi_{\rho\sigma}
\Big).
\label{eq:polsumresult}
\end{equation}

The difference current is symmetric and satisfies
$q_{\mu}\Delta J_{\alpha\beta}^{\mu\nu}=0$ by
Eq.~\eqref{eq:diffward}. Consequently, every correction to
$\Pi_{\mu\nu}=\eta_{\mu\nu}+\cdots$ contains a factor of $q^\mu$
contracted with the current and therefore vanishes. Thus, within this
contraction each $\Pi_{\mu\nu}$ may be replaced by $\eta_{\mu\nu}$. Since the
leading-soft current in Eq.~\eqref{eq:diffcurrent} is real, we obtain
\begin{equation}
\sum_{h}
\big|\eps^{h*}\!\cdot\!\Delta J\big|^{2}
=\Delta J_{\mu\nu}\Delta J^{\mu\nu}
-\frac{1}{2}\big(\Delta J^{\mu}{}_{\mu}\big)^{2}.
\label{eq:deDonder}
\end{equation}
The trace coefficient $-\tfrac12$, generalized to
$-1/(D-2)$ in $D$ spacetime dimensions, is characteristic of the
physical polarization sum for a massless spin-two field.

Substituting Eq.~\eqref{eq:diffcurrent} and using
$p_i^2=-m_i^2$, we find
\begin{align}
\sum_{h}
\big|\eps^{h*}\!\cdot\!\Delta J\big|^{2}
&=
\sum_{i,j\in\alpha\cup\beta}
\tilde\eta_{i}\tilde\eta_{j}\,
\frac{
(p_{i}\cdot p_{j})^{2}
-\tfrac{1}{2}m_{i}^{2}m_{j}^{2}
}{
(p_{i}\cdot q)(p_{j}\cdot q)
}.
\label{eq:polsumdone}
\end{align}
Neither the individual terms in this double sum nor the covariant form
on the right-hand side of Eq.~\eqref{eq:deDonder} are manifestly
nonnegative. Nevertheless, their equality to the sum of
physical-helicity absolute squares ensures nonnegativity at each
momentum $\bm q$ for the conserved difference current considered here.
Current conservation is essential for this covariant rewriting and its
gauge independence; without it, the replacement
$\Pi_{\mu\nu}\to\eta_{\mu\nu}$ would not be valid.

\subsection{Angular integral}
\label{sec:angular}

We write $q^\mu=\omega n^\mu$, with
$n^\mu=(1,\hat{\bm q})$, so that
\begin{equation}
p_i\cdot q
=\omega(p_i\cdot n)
=-\omega\big(E_i-\bm p_i\cdot\hat{\bm q}\big).
\end{equation}
The two soft denominators in Eq.~\eqref{eq:polsumdone} contribute a
factor of $\omega^{-2}$. When combined with the phase-space measure in
Eq.~\eqref{eq:Gamma1}, this leaves the logarithmic radial measure
$\dd\omega/\omega$. The remaining angular dependence is contained in
the master integral
\begin{equation}
I_{ij}
\equiv
\int\!\frac{\dd\Omega_{\hat q}}{4\pi}\,
\frac{(p_i\cdot p_j)^2-\tfrac{1}{2}m_i^2m_j^2}
{(p_i\cdot n)(p_j\cdot n)} .
\label{eq:Iij}
\end{equation}

To evaluate this integral, we combine the two denominators using the
Feynman-parameter identity
\begin{equation}
\frac{1}{AB}
=\int_0^1\!\dd x\,
\frac{1}{[xA+(1-x)B]^2},
\end{equation}
with $A=p_i\cdot n$ and $B=p_j\cdot n$. We also use the elementary
solid-angle identity
\begin{equation}
\int\!\frac{\dd\Omega_{\hat q}}{4\pi}\,
\frac{1}{(P\cdot n)^2}
=-\frac{1}{P^2},
\label{eq:solidangle}
\end{equation}
which holds for a future-directed timelike vector $P^\mu$. Defining
$P_x^\mu\equiv xp_i^\mu+(1-x)p_j^\mu$, we obtain
\begin{equation}
I_{ij}
=\Big[(p_i\cdot p_j)^2-\tfrac{1}{2}m_i^2m_j^2\Big]
\int_0^1\!\frac{\dd x}{-P_x^2}.
\end{equation}

For two future-directed timelike momenta, we introduce their invariant
relative Lorentz factor and relative speed through
\[
-p_i\cdot p_j=m_i m_j\gamma_{ij},
\qquad
\gamma_{ij}=(1-\beta_{ij}^2)^{-1/2},
\]
where
\begin{equation}
\beta_{ij}
\equiv
\sqrt{1-\frac{m_i^2m_j^2}{(p_i\cdot p_j)^2}}.
\label{eq:betaij}
\end{equation}
The denominator of the Feynman-parameter integral can then be written as
\[
-P_x^2
=m_i^2x^2+m_j^2(1-x)^2
+2x(1-x)m_i m_j\gamma_{ij}.
\]
Because $P_x^\mu$ is a convex combination of two future-directed
timelike vectors, it remains future-directed and timelike for
$0\leq x\leq1$. Consequently, $-P_x^2>0$ throughout the integration
interval. For $\beta_{ij}>0$, the integral evaluates to
(Appendix~\ref{app:angular})
\begin{equation}
\int_0^1\!\frac{\dd x}{-P_x^2}
=
\frac{1}{2m_i m_j\gamma_{ij}\beta_{ij}}\,
\ln\frac{1+\beta_{ij}}{1-\beta_{ij}} .
\label{eq:xint}
\end{equation}
This result extends continuously to $\beta_{ij}=0$, for which the
integral equals $1/(m_i m_j)$.

Using
\[
(p_i\cdot p_j)^2-\tfrac12m_i^2m_j^2
=\frac12m_i^2m_j^2\gamma_{ij}^2
\big(1+\beta_{ij}^2\big),
\]
we obtain the closed-form pair contribution
\begin{equation}
I_{ij}
=\frac{m_i m_j}{4}\,f(\beta_{ij}),
\qquad
f(\beta)
\equiv
\frac{1+\beta^2}{\beta\sqrt{1-\beta^2}}\,
\ln\frac{1+\beta}{1-\beta}.
\label{eq:fdef}
\end{equation}
The limit $\beta\to0$ is smooth:
\begin{equation}
f(\beta)
=2+\frac{11}{3}\beta^2+\frac{63}{20}\beta^4
+\order(\beta^6),
\qquad
f(0)=2.
\label{eq:fexp}
\end{equation}
Thus, whenever the two legs have the same four-velocity,
$I_{ij}=m_i m_j/2$. In particular, for $i=j$ one has
$\beta_{ii}=0$ and $I_{ii}=m_i^2/2$, in agreement with a direct
application of Eq.~\eqref{eq:solidangle}.

For a generic noncollinear pair in the ultrarelativistic limit
$\beta_{ij}\to1$,
\[
m_i m_j f(\beta_{ij})
\sim
4(-p_i\cdot p_j)
\ln\!\left[
\frac{2(-p_i\cdot p_j)}{m_i m_j}
\right].
\]
Individual terms in the pairwise representation therefore contain
logarithms that are singular in the zero-mass limit. In the complete
conserved sum, however, the coefficients of these logarithms cancel as
a consequence of
$\sum_i\tilde\eta_i p_i^\mu=0$. Equivalently, in the
physical-polarization representation, the transverse numerator
$p^\mu p^\nu\eps_{\mu\nu}$ vanishes as $\theta^2$ in the collinear
limit, compensating the behavior $p\cdot q\sim\theta^2$ of the soft
denominator and leaving the soft factor finite. Thus, although
collinear finiteness is not manifest term by term in the covariant
pairwise representation, the complete leading soft-graviton factor
has no independent collinear divergence~\cite{Weinberg:1965nx}. If an
exactly massless limit is required, it should be taken only after the
conserved signed sum has been formed.

\subsection{Frequency integral and main result}
\label{sec:main}

Combining Eqs.~\eqref{eq:Gamma1}, \eqref{eq:polsumdone}, and
\eqref{eq:fdef}, and using
$\dd^3q=\omega^2\dd\omega\,\dd\Omega$, we obtain
\begin{align}
\Gamma_{\alpha\beta}
&=
\frac{\kappa^2}{16(2\pi)^3}
\int_\lambda^\Lambda\!\frac{\dd\omega}{\omega}
\int\!\dd\Omega
\nonumber\\[-2pt]
&\quad\times
\sum_{i,j\in\alpha\cup\beta}
\tilde\eta_i\tilde\eta_j
\frac{(p_i\cdot p_j)^2-\tfrac12m_i^2m_j^2}
{(p_i\cdot n)(p_j\cdot n)}
\nonumber\\[2pt]
&=
\frac{\kappa^2}{16(2\pi)^3}\,4\pi
\sum_{i,j\in\alpha\cup\beta}
\tilde\eta_i\tilde\eta_j
\frac{m_i m_j}{4}\,
f(\beta_{ij})
\ln\frac{\Lambda}{\lambda}
\nonumber\\[2pt]
&=
\frac{G}{4\pi}
\sum_{i,j\in\alpha\cup\beta}
\tilde\eta_i\tilde\eta_j
m_i m_j f(\beta_{ij})
\ln\frac{\Lambda}{\lambda},
\end{align}
where the last equality follows from $\kappa^2=32\pi G$, or
equivalently $\kappa^2/(128\pi^2)=G/(4\pi)$. Here, $\lambda$ and
$\Lambda$ denote, respectively, the lower and upper frequency scales
over which the leading-soft approximation is applied. The double sums
run over ordered pairs and include the diagonal terms $i=j$.

It is useful to define the difference $B$-factor by
\begin{equation}
\Delta B_{\alpha\beta}
\equiv
\frac{G}{2\pi}
\sum_{i,j\in\alpha\cup\beta}
\tilde\eta_i\tilde\eta_j\,
m_i m_j f(\beta_{ij})
\geq0.
\label{eq:DeltaB}
\end{equation}
Although its nonnegativity is not manifest in this signed pairwise
representation, it follows from the physical-helicity sum-of-squares
form in Eq.~\eqref{eq:Gamma1}. The decoherence exponent and the
corresponding suppression of the off-diagonal density-matrix element
are therefore
\begin{equation}
\Gamma_{\alpha\beta}
=\frac{\Delta B_{\alpha\beta}}{2}
\ln\frac{\Lambda}{\lambda},
\qquad
|\rho_{\alpha\beta}|
\propto
\left(\frac{\lambda}{\Lambda}\right)^{
\Delta B_{\alpha\beta}/2}.
\label{eq:mainresult}
\end{equation}
Here, the proportionality sign suppresses factors that are independent
of the infrared and resolution scales.

Equation~\eqref{eq:mainresult} reproduces the leading-infrared
suppression obtained in Ref.~\cite{Carney:2017jut}, whose exponent
corresponds to $\Delta B_{\alpha\beta}/2$ in our conventions.
Equivalently, $\Delta B_{\alpha\beta}$ coincides with Weinberg's
exponent $B$~\cite{Weinberg:1965nx} for the formal transition
$\beta\to\alpha$, with the legs of $\beta$ assigned the incoming sign
and those of $\alpha$ the outgoing sign. Thus, in the present
conventions, the decoherence exponent between the two branches is
one-half of the Weinberg exponent associated with the corresponding
branch-to-branch transition.

For the asymptotic momentum branches considered here, we model the
effect of a finite observation time $T$ by an effective infrared
cutoff $\lambda\sim T^{-1}$. Let $E_{\rm res}$ denote the detector
energy threshold below which soft gravitons remain unobserved and
are traced over, so that $\Lambda\sim E_{\rm res}$. Assuming
$T^{-1}\ll E_{\rm res}\ll E_{\rm hard}$, where $E_{\rm hard}$ is a
characteristic energy scale of the hard process, the vacuum result
gives, at leading-logarithmic accuracy,
\begin{equation}
|\rho_{\alpha\beta}(T)|
\propto
\big(E_{\rm res}T\big)^{-\Delta B_{\alpha\beta}/2}.
\label{eq:logtime}
\end{equation}
Thus, $\Gamma_{\alpha\beta}$ grows logarithmically with $T$, whereas
the coherence decays algebraically when $\Delta B_{\alpha\beta}>0$;
no leading-soft suppression occurs when
$\Delta B_{\alpha\beta}=0$. The identification
$\lambda\sim T^{-1}$ is understood here as a sharp-cutoff resolution
prescription; a complete finite-time prediction requires a specified
source history and detector response.

\section{Nonrelativistic expansion: the quadrupole record}
\label{sec:NR}

We now specialize Eq.~\eqref{eq:DeltaB} to branches whose constituent
momenta are nonrelativistic in a common inertial frame,
$|\bm v_i|\ll1$, retaining all terms through $\order(v^4)$. This
expansion shows how energy--momentum conservation eliminates the
lower-order scalar and vector contributions, leaving a symmetric
trace-free rank-two tensor as the leading nonvanishing term. Full details are given in Appendix~\ref{app:NR}; here we outline the logic.

Writing $p_i^\mu=m_i\gamma_i(1,\bm v_i)$, we define the signed velocity
moments of the difference between the two branches by
\begin{align}
C_0&\equiv\sum_i\tilde\eta_i m_i, &
\bm C_1&\equiv\sum_i\tilde\eta_i m_i\bm v_i,
\nonumber\\
\mu^2&\equiv\sum_i\tilde\eta_i m_i\bm v_i^2, &
V^{ab}&\equiv\sum_i\tilde\eta_i m_i v_i^a v_i^b,
\label{eq:moments}
\end{align}
so that $\mu^2=\delta_{ab}V^{ab}$. Here and below, the sums run over
$i\in\alpha\cup\beta$, with the signs $\tilde\eta_i$ defined in
Eq.~\eqref{eq:diffcurrent}. Despite its notation, $\mu^2$ is a signed
branch difference and need not be nonnegative. We reserve $Q^{ab}$ for
the position-space mass quadrupole used in
Ref.~\cite{Matsui:2026whichpath}.

Because the two branches carry the same total four-momentum, their
energy and momentum differences vanish exactly:
\begin{equation}
\sum_i\tilde\eta_i m_i\gamma_i=0,
\qquad
\sum_i\tilde\eta_i m_i\gamma_i\bm v_i=0 .
\label{eq:conservation}
\end{equation}
Expanding these relations in powers of the velocities gives
\begin{equation}
C_0=-\tfrac12\mu^2+\order(v^4),
\qquad
\bm C_1=\order(v^3).
\label{eq:C0C1}
\end{equation}
Thus, the signed rest-mass difference $C_0$ begins at
$\order(v^2)$ and is fixed at this order by minus the leading
kinetic-energy difference. Similarly, the leading mass-current, or
mass-dipole-rate, difference $\bm C_1$ begins at $\order(v^3)$.

We next substitute the expansion in Eq.~\eqref{eq:fexp} into
Eq.~\eqref{eq:DeltaB}. To retain all terms through $\order(v^4)$,
the invariant relative speed must be expanded as
\begin{align}
\beta_{ij}^{2}
&=|\bm v_i-\bm v_j|^{2}
+2(\bm v_i^2+\bm v_j^2)(\bm v_i\!\cdot\!\bm v_j)
\nonumber\\[-2pt]
&\quad
-\bm v_i^{2}\bm v_j^{2}
-3(\bm v_i\!\cdot\!\bm v_j)^{2}
+\order(v^{6}).
\label{eq:betaexp}
\end{align}
The constant term in $f(\beta_{ij})$ produces $2C_0^2$, which begins
at $\order(v^4)$ by Eq.~\eqref{eq:C0C1}. The leading
$\order(\beta_{ij}^2)$ term gives
\[
\frac{11}{3}
\left(2C_0\mu^2-2|\bm C_1|^2\right).
\]
Here, the $C_0\mu^2$ contribution begins at $\order(v^4)$, whereas
the $|\bm C_1|^2$ contribution begins at $\order(v^6)$. Consequently,
the nominal $\order(v^0)$ and $\order(v^2)$ contributions vanish at
their respective orders once the exact conservation laws are imposed.
This cancellation is the leading-soft counterpart of the absence of
monopole and dipole gravitational radiation.

Collecting all contributions through $\order(v^4)$, including those
arising from the higher-order terms in Eq.~\eqref{eq:betaexp}, gives
(Appendix~\ref{app:NR})
\begin{equation}
\Delta B_{\alpha\beta}
=\frac{G}{2\pi}
\left[
\frac{8}{5}V^{ab}V^{ab}
-\frac{8}{15}(\mu^2)^2
\right]
+\order(v^6).
\end{equation}
Decomposing
\[
V^{ab}
=V_{\rm TF}^{ab}+\tfrac13\delta^{ab}\mu^2
\]
into its trace-free and trace parts, we have
\[
V^{ab}V^{ab}
=V_{\rm TF}^{ab}V_{\rm TF}^{ab}
+\tfrac13(\mu^2)^2.
\]
The trace contribution therefore cancels, leaving
\begin{equation}
\begin{aligned}
\Delta B_{\alpha\beta}
&=\frac{4G}{5\pi}\,
V_{\rm TF}^{ab}V_{\rm TF}^{ab}
+\order(v^6),
\\
V_{\rm TF}^{ab}
&=\sum_{i\in\alpha\cup\beta}
\tilde\eta_i m_i
\left(
v_i^a v_i^b
-\tfrac13\delta^{ab}\bm v_i^2
\right).
\end{aligned}
\label{eq:NRfinal}
\end{equation}
Thus, at the leading nonvanishing order in the nonrelativistic
expansion, the distinguishability recorded in the soft-graviton cloud
depends only on the difference between the trace-free second velocity
moments of the two branches. In this precise sense, the leading
soft-graviton which-path record is quadrupolar. This conclusion does
not require the particle contents of the two branches to coincide,
provided that all hard constituents and recoiling degrees of freedom
are included, remain within the nonrelativistic regime, and satisfy
exact energy--momentum conservation. For asymptotically inertial branches, 
the trace-free Newtonian mass
quadrupole
$Q_X^{ab}=[\sum_{i\in X}m_i x_i^a x_i^b]_{\rm TF}$ satisfies
\begin{equation}
V_{\rm TF}^{ab}
=\frac12\left(\ddot Q_\alpha^{ab}-\ddot Q_\beta^{ab}\right)
\end{equation}
at leading order, confirming the quadrupolar character
of Eq.~\eqref{eq:NRfinal}.

Within the same sharp-cutoff resolution prescription, using
Eq.~\eqref{eq:mainresult} with $\lambda\sim T^{-1}$ and
$\Lambda\sim E_{\rm res}$ gives the corresponding vacuum decoherence
exponent
\begin{align}
\Gamma_{\alpha\beta}
&\simeq
\frac{2G}{5\pi}\,
V_{\rm TF}^{ab}V_{\rm TF}^{ab}
\ln\!\big(E_{\rm res}T\big)
\nonumber\\[-2pt]
&\sim
\frac{2}{5\pi}
\left(\frac{m}{\Mpl}\right)^2
v^4\ln\!\big(E_{\rm res}T\big),
\label{eq:NRGamma}
\end{align}
where the second line assumes
$V_{\rm TF}^{ab}V_{\rm TF}^{ab}\sim m^2v^4$. Here, $m$ denotes the
characteristic mass participating in the branch difference, $v$ is
measured in units of the speed of light, and
\[
\Mpl=\sqrt{\hbar c/G}
\simeq\SI{2.18e-8}{\kilogram}
\simeq\SI{22}{\micro\gram}
\]
is the unreduced Planck mass.

For $m\ll\Mpl$ and $v\ll1$, this leading-soft vacuum contribution is
therefore strongly suppressed by both $(m/\Mpl)^2$ and $v^4$. As the
relevant mass scale approaches $\Mpl$, the mass suppression becomes
less severe; nevertheless, the velocity factor, the tensorial
difference between the branches, and the logarithmic observation-time
factor remain essential in determining the magnitude of the effect.

\section{Decoherence in squeezed graviton background}
\label{sec:squeezed}

Within the leading-soft approximation, the gravitational field
conditioned on a hard branch $X$ evolves through the displacement
$\hat D[\A_X]$, together with the branch phase
$\ee^{\ii\Phi_X}$. Assuming that the hard and graviton sectors are
initially uncorrelated, the corresponding off-diagonal element of the
hard reduced density matrix becomes
\begin{align}
\begin{split}
\rho_{\alpha\beta}^{\rm out}
&=
\rho_{\alpha\beta}^{\rm in}\,
\ee^{\ii(\Phi_\alpha-\Phi_\beta)}
\mathcal C_{\alpha\beta}[\rho_g], \\
\mathcal C_{\alpha\beta}[\rho_g]
&\equiv
\Tr_g\!\left[
\rho_g\,
\hat D^\dagger[\A_\beta]\hat D[\A_\alpha]
\right]
=
\ee^{\ii\Theta_{\alpha\beta}}\,
\chi_{\rho_g}[\Delta\A].
\label{eq:chargen}
\end{split}
\end{align}
Here, $\Delta\A\equiv\A_\alpha-\A_\beta$, and the displacement
composition law gives
\begin{align}
\begin{split}
\hat D^\dagger[\A_\beta]\hat D[\A_\alpha]
&=
\ee^{\ii\Theta_{\alpha\beta}}\hat D[\Delta\A], \\
\Theta_{\alpha\beta}
&=
\operatorname{Im}
\sum_h\int\!\frac{\dd^3q}{(2\pi)^3}\,
\A_h^\alpha(\bm q)\A_h^{\beta*}(\bm q).
\end{split}
\end{align}
The functional
\begin{equation}
\chi_{\rho_g}[\xi]
\equiv
\Tr_g\!\left(\rho_g\hat D[\xi]\right)
\end{equation}
is the symmetrically ordered Weyl characteristic functional of the
initial graviton state. Thus, the environmental contribution to the
branch coherence samples this functional at the relative displacement
$\Delta\A$. For a general state, $\chi_{\rho_g}$ may be complex: its
modulus determines the visibility, while its argument contributes an
additional state-dependent relative phase. We therefore define
\begin{equation}
\Gamma_{\alpha\beta}[\rho_g]
\equiv
-\ln\big|\chi_{\rho_g}[\Delta\A]\big|.
\end{equation}
For the graviton vacuum,
\begin{equation}
\chi_0[\xi]
=
\exp\!\left[
-\frac12\sum_h\int\!\frac{\dd^3q}{(2\pi)^3}
|\xi_h(\bm q)|^2
\right],
\end{equation}
which reproduces Eq.~\eqref{eq:Gammadef}. We first use independent
single-mode squeezing as an illustrative prototype and then turn to
the two-mode squeezing relevant to cosmological gravitons
~\cite{Grishchuk:1990bj,Albrecht:1992kf,Parikh:2020kfh,
Parikh:2020fhy,Kanno:2021gpt,Matsui:2025sqg}.

\subsection{Single-mode squeezing}
\label{sec:singlemode}

Consider a product of independently squeezed modes,
$\rho_g=\hat S[z]|0\rangle\langle0|\hat S^\dagger[z]$, with
\begin{equation}
\hat S[z]
=
\exp\!\Bigg[
\frac12\sum_h\int\!\frac{\dd^3q}{(2\pi)^3}
\left(
z_q\,a_h^{\dagger2}(\bm q)
-z_q^*\,a_h^2(\bm q)
\right)
\Bigg],
\end{equation}
where $z_q=r_q\ee^{\ii\varphi_q}$.

The continuous-mode expressions may be understood as the continuum
limit of a wave-packet mode basis. With the
convention adopted here, the Bogoliubov transformation is
\begin{equation}
\hat S^\dagger a_h(\bm q)\hat S
=
a_h(\bm q)\cosh r_q
+\ee^{\ii\varphi_q}a_h^\dagger(\bm q)\sinh r_q .
\end{equation}
Consequently,
\begin{align}
\begin{split}
\hat S^\dagger\hat D[\xi]\hat S
&=
\hat D[\xi'], \\
\xi'_h(\bm q)
&=
\xi_h(\bm q)\cosh r_q
-\ee^{\ii\varphi_q}\xi_h^*(\bm q)\sinh r_q .
\end{split}
\end{align}

The characteristic functional of the squeezed vacuum is therefore
\begin{align}
\begin{split}
\chi_{\rm sq}[\xi]
&=
\exp\!\Bigg[
-\frac12\sum_h\int\!\frac{\dd^3q}{(2\pi)^3}
|\xi'_h(\bm q)|^2
\Bigg],
\\
|\xi'_h|^2
&=
|\xi_h|^2\cosh 2r_q
-\sinh 2r_q\,
\operatorname{Re}\!\left(
\ee^{\ii\varphi_q}\xi_h^{*2}
\right).
\end{split}
\end{align}
It follows that
\begin{equation}
\begin{aligned}
\Gamma^{\rm sq}_{\alpha\beta}
=
\frac12\sum_h\int\!\frac{\dd^3q}{(2\pi)^3}
\Big[
&\cosh 2r_q\,|\Delta\A_h(\bm q)|^2
\\[-2pt]
&-\sinh 2r_q\,
\operatorname{Re}\!\left(
\ee^{\ii\varphi_q}
[\Delta\A_h^*(\bm q)]^2
\right)
\Big].
\end{aligned}
\label{eq:Gammasq}
\end{equation}
Writing
$\Delta\A_h(\bm q)=|\Delta\A_h(\bm q)|
\ee^{\ii\theta_h(\bm q)}$, each mode is weighted relative to its vacuum
contribution by
\begin{align}
\begin{split}
g_h(\bm q)
=\cosh 2r_q &-\cos\!\big(\varphi_q-2\theta_h(\bm q)\big)\sinh 2r_q, \\
\ee^{-2r_q}&\leq g_h(\bm q)\leq\ee^{2r_q}.
\end{split}
\end{align}
Thus, squeezing can suppress or enhance decoherence depending on the
relative quadrature phase.

\subsection{Two-mode squeezing and the relic graviton background}
\label{sec:twomode}

In the standard decomposition of tensor perturbations on an FLRW
background, the cosmological expansion correlates modes of opposite wave vector,
$\bm{k}$ and $-\bm{k}$. A homogeneous and isotropic primordial graviton state that
evolves from an initial vacuum can therefore be written as a product of two-mode
squeezed vacua over independent mode pairs and polarizations,
\begin{align}
|Z\rangle&=\hat S_2[z]|0\rangle,
\nonumber\\
\hat S_2[z]
&=
\exp\!\Bigg\{
\frac12\sum_h\int\!\frac{\dd^3q}{(2\pi)^3}
\Big[
z_q\,a_h^\dagger(\bm q)a_h^\dagger(-\bm q) \nonumber\\[-2pt]
&-z_q^*\,a_h(\bm q)a_h(-\bm q)
\Big]
\Bigg\}.
\end{align}
Here, $z_q=r_q\ee^{\ii\varphi_q}$ and
$z_{-\bm q}=z_{\bm q}$. Because the integral extends over the full
momentum space, the factor $1/2$ ensures that each unordered pair
$\{\bm q,-\bm q\}$ is counted only once. Equivalently, one may
integrate over one representative of each pair and omit this factor.
The two-mode representation in the traveling-wave basis is equivalent
to single-mode squeezing of the corresponding standing-wave modes.

With these conventions,
\begin{equation}
\hat S_2^\dagger a_h(\bm q)\hat S_2
=
a_h(\bm q)\cosh r_q
+\ee^{\ii\varphi_q}a_h^\dagger(-\bm q)\sinh r_q ,
\end{equation}
and the displacement argument transforms as
\begin{equation}
\xi'_h(\bm q)
=
\xi_h(\bm q)\cosh r_q
-\ee^{\ii\varphi_q}\xi_h^*(-\bm q)\sinh r_q .
\end{equation}
Using the vacuum characteristic functional of the transformed
displacement gives
\begin{align}
\Gamma^{\rm 2sq}_{\alpha\beta}
&=
\frac12\sum_h\int\!\frac{\dd^3q}{(2\pi)^3}
\Bigg\{
\cosh 2r_q\,|\Delta\A_h(\bm q)|^2
\nonumber\\[-2pt]
&\hspace{35pt}
-\sinh 2r_q\,
\operatorname{Re}\!\left[
\ee^{-\ii\varphi_q}
\Delta\A_h(\bm q)\Delta\A_h(-\bm q)
\right]
\Bigg\}, 
\label{eq:Gamma2sq}
\end{align}
which is valid at leading soft order without making a
nonrelativistic approximation.

We next consider its nonrelativistic limit. Let
$q_\pm^\mu=\omega(1,\pm\hat{\bm q})$. For
$p_i^\mu=m_i\gamma_i(1,\bm v_i)$,
\begin{equation}
p_i\cdot q_\pm
=
-m_i\gamma_i\omega
\big(1\mp\bm v_i\cdot\hat{\bm q}\big).
\end{equation}
Because the transverse--traceless polarization tensor has only spatial
components in this frame, the soft displacement begins at
$\order(v^2)$. Denoting this leading contribution by
$\Delta\A_h^{(2)}$, one finds
\begin{equation}
\Delta\A_h^{(2)}(\bm q)
=
-\frac{\kappa}{2\sqrt{2}\,\omega^{3/2}}\,
\eps^{h*}_{ab}(\bm q)V^{ab}.
\end{equation}
The parity convention in Eq.~\eqref{eq:parity}, together with the
reality of $V^{ab}$, implies
\[
\Delta\A_h^{(2)}(-\bm q)
=
\Delta\A_h^{(2)*}(\bm q).
\]
More generally,
\begin{equation}
\Delta\A_h(-\bm q)
=
\Delta\A_h^*(\bm q)+\order(v^3),
\label{eq:reality}
\end{equation}
where the correction is of relative order $v$ with respect to the
generic leading contribution $\Delta\A_h=\order(v^2)$. It originates
from the direction-odd terms in the expansion of
$(p_i\cdot q_\pm)^{-1}$.

At the leading nonvanishing nonrelativistic order, Eq.~\eqref{eq:Gamma2sq}
therefore reduces to
\begin{align}
\Gamma^{\rm 2sq}_{\alpha\beta}
&\simeq
\frac12\sum_h\int\!\frac{\dd^3q}{(2\pi)^3}
\Big[
\cosh 2r_q-\cos\varphi_q\,\sinh 2r_q
\Big]
\nonumber\\[-2pt]
&\hspace{65pt}\times
|\Delta\A_h^{(2)}(\bm q)|^2 .
\label{eq:Gamma2sqfinal}
\end{align}
For each opposite-momentum pair,
\[
\ee^{-2r_q}
\leq
\cosh 2r_q-\cos\varphi_q\,\sinh 2r_q
\leq
\ee^{2r_q}.
\]
Thus, in this limit, the relevant quadrature is selected by the
squeezing, or standing-wave, phase $\varphi_q$, rather than by the
phase of the individual traveling-wave soft factor.

In the usual canonical description, $r_q$ grows while a mode is
outside the horizon and remains parametrically large after reentry.
For well-separated horizon-exit and reentry epochs,
\[
r_q
\sim
\ln\!\frac{a_{\rm re}(q)}{a_{\rm exit}(q)},
\]
up to background- and convention-dependent corrections
~\cite{Grishchuk:1990bj,Albrecht:1992kf}. For subhorizon relic modes,
the squeezing phase generally varies rapidly with frequency. If the
response-weighted band of width $\Delta\omega$ satisfies
$|\partial_\omega\varphi_q|\Delta\omega\gg1$, while the remaining
factors vary slowly over the corresponding oscillation scale, the
phase-sensitive contribution is suppressed by frequency coarse
graining. Recent analyses have likewise emphasized that squeezing
defined in global source modes need not remain phase resolved in the
wave-packet modes accessible to a local detector; mode reduction,
angular acceptance, and phase incoherence can make the observed state
effectively thermal~\cite{Miyauchi:2026cnt}. One then
obtains
\begin{equation}
\Gamma^{\rm 2sq,av}_{\alpha\beta}
\simeq
\frac12\sum_h\int\!\frac{\dd^3q}{(2\pi)^3}
(1+2n_q)|\Delta\A_h(\bm q)|^2,
\label{eq:relicavg}
\end{equation}
where $n_q=\sinh^2r_q $. Equation~\eqref{eq:Gamma2sq} should be retained for a phase-resolved
state. Equation~\eqref{eq:Gamma2sqfinal} is its leading
nonrelativistic form, whereas the stochastic-background estimate below
assumes the frequency coarse graining used in
Eq.~\eqref{eq:relicavg}.

\subsection{Frequency integral and infrared scaling}
\label{sec:powerlaw}

After frequency coarse graining, assume that the background is
statistically isotropic and unpolarized, with the same occupation
number $n_\omega$ for both helicities. Equation~\eqref{eq:relicavg}
then becomes
\begin{equation}
\Gamma^{\rm sq,av}_{\alpha\beta}
=
\frac{\Delta B_{\alpha\beta}}{2}
\int_\lambda^\Lambda\frac{\dd\omega}{\omega}
\big(1+2n_\omega\big).
\label{eq:sqmaster}
\end{equation}
The normal-ordered energy density of the two graviton helicities is
\begin{equation}
\rho_{\rm gw}
=
\sum_h\int\!\frac{\dd^3q}{(2\pi)^3}\,
\omega n_\omega
=
\frac{1}{\pi^2}\int\!\dd\omega\,\omega^3n_\omega .
\end{equation}
Defining
\[
\Omega_{\rm gw}(\omega)
\equiv
\frac{1}{\varepsilon_{c,0}}
\frac{\dd\rho_{\rm gw}}{\dd\ln\omega},
\]
where $\varepsilon_{c,0}$ is the present critical energy density, gives
\begin{equation}
n_\omega
=
\frac{\pi^2\varepsilon_{c,0}}{\omega^4}\,
\Omega_{\rm gw}(\omega).
\label{eq:noccup}
\end{equation}
Consequently,
\begin{equation}
\Gamma^{\rm sq,av}_{\alpha\beta}
=
\frac{\Delta B_{\alpha\beta}}{2}
\ln\frac{\Lambda}{\lambda}
+
\pi^2\Delta B_{\alpha\beta}\varepsilon_{c,0}
\int_\lambda^\Lambda
\frac{\dd\omega}{\omega^5}\,
\Omega_{\rm gw}(\omega).
\end{equation}

For a power-law spectrum $\Omega_{\rm gw}\propto\omega^{n_t}$, the
occupation-dependent contribution is infrared dominated when $n_t<4$.
If the integration band lies within an approximately flat
inflationary plateau,
$\Omega_{\rm gw}(\omega)\simeq\Omega_{\rm gw}$, then
\begin{equation}
\Gamma^{\rm sq,av}_{\alpha\beta}
=
\frac{\Delta B_{\alpha\beta}}{2}
\ln\frac{\Lambda}{\lambda}
+
\frac{\pi^2}{4}\Delta B_{\alpha\beta}
\varepsilon_{c,0}\Omega_{\rm gw}
\left(\lambda^{-4}-\Lambda^{-4}\right).
\end{equation}
Within the sharp-cutoff prescription $\lambda=T^{-1}$ and for
$\Lambda T\gg1$, this reduces to
\begin{equation}
\Gamma^{\rm sq,av}_{\alpha\beta}(T)
\simeq
\frac{\Delta B_{\alpha\beta}}{2}\ln(\Lambda T)
+
\frac{\pi^2}{4}\Delta B_{\alpha\beta}
\varepsilon_{c,0}\Omega_{\rm gw}T^4 .
\label{eq:T4}
\end{equation}

The first term is the vacuum contribution, whereas the second arises
from the state-dependent occupation of the graviton modes and reduces
to classical stochastic-noise dephasing when $n_\omega\gg1$. This
structure is consistent with influence-functional and master-equation
treatments of graviton-induced noise and decoherence
~\cite{Blencowe:2012mp,Kanno:2020usf,Parikh:2020nrd,
Toros:2020krn}.

The $T^4$ estimate follows from the sharp-cutoff identification
$\lambda=T^{-1}$ and is not established here as a universal
finite-time law. Its applicability requires the phase coarse graining
to remain valid near $\omega\sim T^{-1}$, the infrared end of the
integration band to lie within the approximately flat, subhorizon
portion of the spectrum, and
$T^{-1}\leq\Lambda\ll E_{\rm hard}$. A low-frequency turnover or a
physical infrared cutoff changes the late-time behavior. The
numerical coefficient of the power-law term also depends on the
specified source switching and detector response.

\section{Discussion}
\label{sec:discussion}

Using branch-conditioned coherent states, we have recovered the leading
infrared suppression of hard-branch coherence in the graviton
vacuum~\cite{Carney:2017jut} and extended the same framework to squeezed
graviton states. The vacuum exponent is one half of the squared norm of
the relative radiation displacement, $\Gamma_{\alpha\beta}=N_\Delta/2$.
For nonrelativistic branches with the same total four-momentum,
energy--momentum conservation removes the monopole and dipole
contributions, leaving the trace-free second velocity moment in
Eq.~\eqref{eq:NRfinal}. At this order the initial graviton state enters
through its characteristic function: phase-resolved squeezing may
suppress or enhance decoherence, whereas phase averaging produces the
occupation-number enhancement in Eq.~\eqref{eq:relicavg}. The $T^4$
behavior in Eq.~\eqref{eq:T4} is specific to the flat-spectrum,
sharp-cutoff estimate and is not a universal finite-time law.

Infrared-finite scattering amplitudes and reduced-state coherence are
different observables. Coherent dressings reorganize the soft sector,
but tracing over unresolved modes can still suppress coherence;
comparisons therefore require the same physical state, observable, and
hard--soft partition~\cite{Carney:2017oxp,Prabhu:2022zcr,Prabhu:2024lmg}.

Equation~\eqref{eq:NRGamma} shows that the vacuum contribution is
strongly suppressed by $(m/\Mpl)^2(v/c)^4$ for ordinary laboratory
matter waves. A large graviton occupation can enhance the leading-soft
contribution, but its observability depends on the accessible frequency
band, phase coherence, source history, and detector response. The
present asymptotic-branch analysis is complementary to horizon-induced
decoherence, where a time-dependent difference current can produce
different long-time scaling~\cite{Danielson:2022tdw,Danielson:2022sga}.

Our calculation is limited to leading order in the soft expansion and
in $\kappa$. In occupied or squeezed states, state weighting can amplify
subleading soft terms, so their neglect must be checked over the
frequencies that dominate the integral. Realistic wave packets also
require averaging the branch-difference functional over the momentum
distribution. Finally, $\lambda\sim T^{-1}$ is a resolution
prescription; a quantitative finite-time prediction, especially for
the occupation-dependent term, requires a specified switching history
and detector response.

\begin{acknowledgments}
This work is supported by JSPS KAKENHI Grant No. JP23K13100.
\end{acknowledgments}

\appendix

\section{Graviton polarization sum}
\label{app:polsum}

Let
$q^\mu=\omega(1,\hat{\bm q})$ and
$\bar q^\mu=\omega(1,-\hat{\bm q})$, so that
$q\cdot\bar q=-2\omega^2$. Choose helicity vectors
$\eps_\mu^s(\bm q)$, with $s=\pm$, satisfying
\begin{equation}
\begin{aligned}
q\cdot\eps^s&=\bar q\cdot\eps^s=0,\\
\eps_\mu^s\eps^{s'*\mu}&=\delta^{ss'}.
\end{aligned}
\label{eq:apphelicitynorm}
\end{equation}
Their completeness relation on the two-dimensional subspace
orthogonal to $q^\mu$ and $\bar q^\mu$ is
\begin{equation}
\sum_{s=\pm}\eps_\mu^s\eps_\nu^{s*}
=\Pi_{\mu\nu},
\qquad
\Pi_{\mu\nu}
=
\eta_{\mu\nu}
-\frac{q_\mu\bar q_\nu+\bar q_\mu q_\nu}
{q\cdot\bar q}.
\label{eq:photoncompl}
\end{equation}

The graviton polarization tensors are
$\eps_{\mu\nu}^{\pm2}=\eps_\mu^\pm\eps_\nu^\pm$ and satisfy the
normalization, transversality, and tracelessness conditions in
Eq.~\eqref{eq:polnorm}. To derive their completeness relation,
introduce real orthonormal transverse vectors $e_1^\mu$ and $e_2^\mu$
such that
\begin{equation}
\begin{aligned}
\Pi_{\mu\nu}
&=e_{1\mu}e_{1\nu}+e_{2\mu}e_{2\nu},\\
\eps_\mu^\pm
&=\frac{e_{1\mu}\pm\ii e_{2\mu}}{\sqrt2}.
\end{aligned}
\label{eq:apphelicitybasis}
\end{equation}
Direct expansion then gives
\begin{equation}
\sum_{h=\pm2}
\eps^h_{\mu\nu}\eps^{h*}_{\rho\sigma}
=
\frac12
\left(
\Pi_{\mu\rho}\Pi_{\nu\sigma}
+\Pi_{\mu\sigma}\Pi_{\nu\rho}
-\Pi_{\mu\nu}\Pi_{\rho\sigma}
\right).
\label{eq:apppolsum}
\end{equation}
The right-hand side is transverse and traceless. Its full contraction
is
$\tfrac12(2^2+2-2)=2$, as required for the two physical graviton
helicities.

Finally, let $J^{\mu\nu}$ be a symmetric, possibly complex current
satisfying $q_\mu J^{\mu\nu}=0$. In the bilinear contraction of
Eq.~\eqref{eq:apppolsum}, every term containing
$\Pi_{\mu\nu}-\eta_{\mu\nu}$ vanishes because it contains at least one
factor of $q^\mu$ contracted with a current. Hence
\begin{equation}
\sum_h
\left|\eps_{\mu\nu}^{h*}J^{\mu\nu}\right|^2
=
J_{\mu\nu}J^{*\mu\nu}
-\frac12\left|J^\mu{}_\mu\right|^2.
\end{equation}
For the real eikonal current used in the main text, this reduces to
Eq.~\eqref{eq:deDonder}.

\section{The master angular integral}
\label{app:angular}

We derive Eqs.~\eqref{eq:solidangle}--\eqref{eq:xint}. Let
$P^\mu=(P^0,\bm P)$ be future-directed and timelike. Choosing the polar
axis along $\bm P$, and using the metric signature $(-,+,+,+)$, gives
\begin{align}
\int\!\frac{\dd\Omega}{4\pi}\frac{1}{(P\cdot n)^2}
&=
\frac12\int_{-1}^{1}
\frac{\dd c}{(P^0-|\bm P|c)^2}
\nonumber\\
&=
\frac{1}{2|\bm P|}
\left(
\frac{1}{P^0-|\bm P|}
-\frac{1}{P^0+|\bm P|}
\right)
=-\frac{1}{P^2}.
\end{align}
This establishes Eq.~\eqref{eq:solidangle}.

We next apply
\begin{equation}
\frac{1}{AB}
=
\int_0^1\!\frac{\dd x}
{[xA+(1-x)B]^2}
\label{eq:appfeynman}
\end{equation}
with $A=p_i\cdot n$ and $B=p_j\cdot n$. Defining
$P_x^\mu=xp_i^\mu+(1-x)p_j^\mu$, we obtain
\begin{equation}
\int\!\frac{\dd\Omega}{4\pi}
\frac{1}{(p_i\cdot n)(p_j\cdot n)}
=
\int_0^1\!\frac{\dd x}{-P_x^2}
\equiv J.
\end{equation}
Because $P_x^\mu$ is a convex combination of future-directed timelike
vectors, it remains timelike for $0\leq x\leq1$, and hence
$-P_x^2>0$ throughout the integration interval.

Introduce the relative rapidity $\zeta_{ij}$ through
\begin{equation}
\begin{aligned}
\gamma_{ij}&=\cosh\zeta_{ij},
&\beta_{ij}&=\tanh\zeta_{ij},\\
\gamma_{ij}\beta_{ij}&=\sinh\zeta_{ij}.
\end{aligned}
\label{eq:apprapidity}
\end{equation}
Using $-p_i\cdot p_j=m_i m_j\gamma_{ij}$, we have
\begin{equation}
\begin{aligned}
-P_x^2
={}&m_i^2x^2+m_j^2(1-x)^2
\\
&+2m_i m_j\gamma_{ij}x(1-x).
\end{aligned}
\label{eq:appPxquadratic}
\end{equation}
The substitution $u=x/(1-x)$ gives
\begin{align}
J
&=
\int_0^\infty
\frac{\dd u}
{(m_i u+m_j\ee^{\zeta_{ij}})
 (m_i u+m_j\ee^{-\zeta_{ij}})}
\nonumber\\
&=
\frac{\zeta_{ij}}{m_i m_j\sinh\zeta_{ij}}
=
\frac{1}{2m_i m_j\gamma_{ij}\beta_{ij}}
\ln\frac{1+\beta_{ij}}{1-\beta_{ij}},
\end{align}
which is Eq.~\eqref{eq:xint}. This expression extends continuously to
$\beta_{ij}=0$, where $J=1/(m_i m_j)$.

Finally,
\begin{equation}
\begin{aligned}
(p_i\cdot p_j)^2-\frac12m_i^2m_j^2
={}&\frac12m_i^2m_j^2\gamma_{ij}^2
\\
&\times(1+\beta_{ij}^2),
\end{aligned}
\label{eq:appangularnumerator}
\end{equation}
and therefore
\begin{equation}
\begin{aligned}
I_{ij}
={}&\frac{m_i m_j}{4}
\frac{1+\beta_{ij}^2}
{\beta_{ij}\sqrt{1-\beta_{ij}^2}}
\\
&\times
\ln\frac{1+\beta_{ij}}{1-\beta_{ij}},
\end{aligned}
\label{eq:appIijresult}
\end{equation}
in agreement with Eq.~\eqref{eq:fdef}. In particular,
$I_{ij}=m_i m_j/2$ when $\beta_{ij}=0$, and hence
$I_{ii}=m_i^2/2$. Using
\begin{equation}
f(\beta)
=
2(1+\beta^2)(1-\beta^2)^{-1/2}
\frac{\operatorname{arctanh}\beta}{\beta},
\label{eq:appfalt}
\end{equation}
we obtain
\begin{equation}
\begin{aligned}
f(\beta)
={}&2+\frac{11}{3}\beta^2
+\frac{63}{20}\beta^4
\\
&+\order(\beta^6),
\end{aligned}
\label{eq:appfseries}
\end{equation}
which reproduces Eq.~\eqref{eq:fexp}.

\section{Displacement and squeeze operator identities}
\label{app:squeeze}

\subsection{Displacements and coherent overlaps}

For one normalized oscillator mode, let
\begin{equation}
\hat D(\xi)=\exp(\xi a^\dagger-\xi^*a).
\label{eq:appDdef}
\end{equation}
The Baker--Campbell--Hausdorff formula gives
\begin{equation}
\begin{aligned}
\hat D(\xi_1)\hat D(\xi_2)
&=
\ee^{\ii\operatorname{Im}(\xi_1\xi_2^*)}
\hat D(\xi_1+\xi_2),\\
\hat D^\dagger(\xi)&=\hat D(-\xi).
\end{aligned}
\label{eq:appDcomposition}
\end{equation}
For regulated, square-integrable multimode profiles, define
\begin{equation}
\|\xi\|^2
\equiv
\sum_h\int\!\frac{\dd^3q}{(2\pi)^3}
|\xi_h(\bm q)|^2.
\end{equation}
It follows that
\begin{equation}
\begin{aligned}
\hat D^\dagger[\A_\beta]\hat D[\A_\alpha]
&=
\ee^{\ii\Theta_{\alpha\beta}}
\hat D[\Delta\A], \\
\Theta_{\alpha\beta}
&=
\operatorname{Im}
\sum_h\int\!\frac{\dd^3q}{(2\pi)^3}
\A^\alpha_h(\bm q)\A^{\beta*}_h(\bm q).
\end{aligned}
\label{eq:appDrelative}
\end{equation}
Using
$\langle0|\hat D[\xi]|0\rangle=\exp[-\|\xi\|^2/2]$, we obtain
\begin{equation}
\langle\gamma_\beta|\gamma_\alpha\rangle
=
\exp\!\left[
\ii\Theta_{\alpha\beta}
-\frac12\|\Delta\A\|^2
\right],
\end{equation}
and hence Eq.~\eqref{eq:Gammadef}.

\subsection{Single-mode squeezing}

For a normalized discrete or wave-packet mode, consider
\begin{equation}
\hat S(z)
=
\exp\!\left[
\frac12
\left(
z\,a^{\dagger2}-z^*a^2
\right)
\right],
\qquad
z=r\ee^{\ii\varphi}.
\end{equation}
With this convention,
\begin{equation}
\hat S^\dagger(z)a\hat S(z)
=
a\cosh r+a^\dagger\ee^{\ii\varphi}\sinh r.
\end{equation}
Consequently,
\begin{align}
\hat S^\dagger\hat D(\xi)\hat S
&=\hat D(\xi'),
&
\xi'
&=\xi\cosh r-\xi^*\ee^{\ii\varphi}\sinh r.
\end{align}
The characteristic function of the squeezed vacuum is therefore
\begin{align}
\chi_{\rm sq}(\xi)
&=
\langle0|\hat S^\dagger\hat D(\xi)\hat S|0\rangle
=
\exp\!\left(-\frac12|\xi'|^2\right),
\\
|\xi'|^2
&=
|\xi|^2\cosh2r
-\sinh2r\,
\operatorname{Re}\!\left(
\ee^{\ii\varphi}\xi^{*2}
\right).
\end{align}
Writing $\xi=|\xi|\ee^{\ii\theta}$ gives
\begin{equation}
|\xi'|^2
=
|\xi|^2
\left[
\cosh2r-\cos(\varphi-2\theta)\sinh2r
\right].
\end{equation}
The multiplicative factor lies between $\ee^{-2r}$ and $\ee^{2r}$.
Its uniform phase average at the level of this quadratic form is
$\cosh2r=1+2\sinh^2r$.

\subsection{Two-mode squeezing}

For one unordered pair
$(a_+,a_-)\equiv(a_h(\bm q),a_h(-\bm q))$, define
\begin{equation}
\hat S_2(z)
=
\exp\!\left[
z\,a_+^\dagger a_-^\dagger-z^*a_+a_-
\right].
\end{equation}
This fixed-pair expression contains no factor of $1/2$. When the
generator is integrated over the full momentum space, the factor
$1/2$ in Sec.~\ref{sec:twomode} prevents each pair
$\{\bm q,-\bm q\}$ from being counted twice. The Bogoliubov
transformation is
\begin{equation}
\hat S_2^\dagger a_\pm\hat S_2
=
a_\pm\cosh r
+a_\mp^\dagger\ee^{\ii\varphi}\sinh r.
\end{equation}
For
\begin{equation}
\hat D(\xi_+,\xi_-)
=
\exp\!\left[
\sum_{\sigma=\pm}
\left(
\xi_\sigma a_\sigma^\dagger
-\xi_\sigma^*a_\sigma
\right)
\right],
\end{equation}
we find
\begin{align}
\hat S_2^\dagger
\hat D(\xi_+,\xi_-)
\hat S_2
&=
\hat D(\xi'_+,\xi'_-),
\nonumber\\
\xi'_\pm
&=
\xi_\pm\cosh r
-\xi_\mp^*\ee^{\ii\varphi}\sinh r.
\end{align}
It follows that
\begin{align}
|\xi'_+|^2+|\xi'_-|^2
&=
\bigl(|\xi_+|^2+|\xi_-|^2\bigr)\cosh2r
\nonumber\\
&\quad
-2\sinh2r\,
\operatorname{Re}\!\left(
\ee^{-\ii\varphi}\xi_+\xi_-
\right).
\end{align}
This gives Eq.~\eqref{eq:Gamma2sq} upon setting
$\xi_\pm=\Delta\A_h(\pm\bm q)$. If
$\xi_-=\xi_+^*$, the interference term becomes
$|\xi_+|^2\cos\varphi$, yielding
Eq.~\eqref{eq:Gamma2sqfinal}.

Using
$2|\xi_+\xi_-|\leq|\xi_+|^2+|\xi_-|^2$, we obtain
\begin{equation}
\ee^{-2r}
\bigl(|\xi_+|^2+|\xi_-|^2\bigr)
\leq
|\xi'_+|^2+|\xi'_-|^2
\leq
\ee^{2r}
\bigl(|\xi_+|^2+|\xi_-|^2\bigr).
\end{equation}
The exponent is therefore nonnegative, consistently with
$|\chi|\leq1$.

\section{Details of the nonrelativistic expansion}
\label{app:NR}

We expand Eq.~\eqref{eq:DeltaB} through $\order(v^4)$. All sums below
run over $i\in\alpha\cup\beta$ with signs $\tilde\eta_i$. In addition
to the moments defined in Eq.~\eqref{eq:moments}, introduce
\begin{equation}
M_4
\equiv
\sum_i\tilde\eta_i m_i(\bm v_i^2)^2,
\qquad
\bm K
\equiv
\sum_i\tilde\eta_i m_i\bm v_i^2\bm v_i.
\end{equation}
Both are signed branch differences. Likewise, $\mu^2$ denotes the
signed second moment defined in Eq.~\eqref{eq:moments}, rather than the
square of a quantity $\mu$; its square will therefore be written
explicitly as $(\mu^2)^2$.

Expanding the exact energy and momentum conservation laws gives
\begin{align}
0
&=
C_0+\frac12\mu^2+\frac38M_4+\order(v^6),
\\
\bm 0
&=
\bm C_1+\frac12\bm K+\order(v^5).
\end{align}
Thus,
\begin{equation}
C_0
=
-\frac12\mu^2-\frac38M_4+\order(v^6),
\qquad
\bm C_1
=
-\frac12\bm K+\order(v^5).
\end{equation}
Only the leading relations
$C_0=-\mu^2/2+\order(v^4)$ and
$\bm C_1=\order(v^3)$ are required below.

Using
$\gamma_{ij}=\gamma_i\gamma_j
(1-\bm v_i\cdot\bm v_j)$, we obtain
\begin{align}
\beta_{ij}^2
&=
1-
\frac{(1-\bm v_i^2)(1-\bm v_j^2)}
{(1-\bm v_i\cdot\bm v_j)^2}
\nonumber\\
&=
|\bm v_i-\bm v_j|^2
+2(\bm v_i^2+\bm v_j^2)
(\bm v_i\cdot\bm v_j)
\nonumber\\
&\quad
-\bm v_i^2\bm v_j^2
-3(\bm v_i\cdot\bm v_j)^2
+\order(v^6),
\end{align}
as stated in Eq.~\eqref{eq:betaexp}. Moreover,
\begin{equation}
\beta_{ij}^4
=
|\bm v_i-\bm v_j|^4+\order(v^6).
\end{equation}

For convenience, define
\begin{equation}
\langle X_{ij}\rangle
\equiv
\sum_{i,j\in\alpha\cup\beta}
\tilde\eta_i\tilde\eta_j
m_i m_j X_{ij}.
\end{equation}
Let $d_{ij}^2\equiv|\bm v_i-\bm v_j|^2$ and write
$\beta_{ij}^2=d_{ij}^2+\delta_{ij}^{(4)}+\order(v^6)$. The required
signed sums are
\begin{align}
\langle2\rangle
&=
2C_0^2
=
\frac12(\mu^2)^2+\order(v^6),
\\
\langle d_{ij}^2\rangle
&=
2C_0\mu^2-2|\bm C_1|^2
=
-(\mu^2)^2+\order(v^6),
\\
\langle\delta_{ij}^{(4)}\rangle
&=
4\bm K\cdot\bm C_1
-(\mu^2)^2
-3V^{ab}V^{ab}
\nonumber\\
&=
-(\mu^2)^2
-3V^{ab}V^{ab}
+\order(v^6),
\\
\langle d_{ij}^4\rangle
&=
2C_0M_4-8\bm K\cdot\bm C_1
+4V^{ab}V^{ab}
+2(\mu^2)^2
\nonumber\\
&=
4V^{ab}V^{ab}
+2(\mu^2)^2
+\order(v^6).
\end{align}
Consequently,
\begin{align}
\langle\beta_{ij}^2\rangle
&=
-2(\mu^2)^2-3V^{ab}V^{ab}
+\order(v^6),
\\
\langle\beta_{ij}^4\rangle
&=
4V^{ab}V^{ab}+2(\mu^2)^2
+\order(v^6).
\end{align}

Substituting
$f(\beta)=2+\tfrac{11}{3}\beta^2
+\tfrac{63}{20}\beta^4+\order(\beta^6)$ gives
\begin{equation}
\begin{aligned}
\Delta B_{\alpha\beta}
={}&\frac{G}{2\pi}\Bigg\{
\frac12(\mu^2)^2
\\[-2pt]
&+\frac{11}{3}
\left[-2(\mu^2)^2-3V^{ab}V^{ab}\right]
\\[-2pt]
&+\frac{63}{20}
\left[4V^{ab}V^{ab}+2(\mu^2)^2\right]
\Bigg\}
\\[-2pt]
&+\order(v^6).
\end{aligned}
\label{eq:appNRcollect}
\end{equation}
Collecting the two tensor structures gives
\begin{equation}
\begin{aligned}
\Delta B_{\alpha\beta}
={}&\frac{G}{2\pi}
\left[
\frac85V^{ab}V^{ab}
-\frac{8}{15}(\mu^2)^2
\right]
\\[-2pt]
&+\order(v^6).
\end{aligned}
\label{eq:appNRbeforeTF}
\end{equation}
Since
\begin{equation}
V^{ab}V^{ab}
=
V_{\rm TF}^{ab}V_{\rm TF}^{ab}
+\frac13(\mu^2)^2,
\end{equation}
the trace terms cancel at this order, yielding
\begin{equation}
\Delta B_{\alpha\beta}
=
\frac{4G}{5\pi}
V_{\rm TF}^{ab}V_{\rm TF}^{ab}
+\order(v^6),
\end{equation}
in agreement with Eq.~\eqref{eq:NRfinal}.

Together with energy--momentum conservation, these cancellations remove
the lower-order scalar and vector contributions, leaving the
difference between the trace-free second velocity moments as the
leading soft record. This conclusion does not require the two branches
to have identical particle content, provided that all relevant hard
and recoiling degrees of freedom are included and remain
nonrelativistic.

\bibliographystyle{JHEP}
\bibliography{References}

\end{document}